\documentclass[aps,prl,showpacs,twocolumn,a4paper,10pt]{revtex4}
\usepackage{graphicx}
\usepackage{amsmath}
\usepackage{amssymb}

\begin{document}

\title{Dipole-Dipole interaction in photonic crystal nanocavity}
\author{Yong-Gang Huang,$^{1,2,3}$ Gengyan Chen,$^{1}$ Chong-Jun Jin,$^{1}$
W. M. Liu,$^{3}$ and Xue-Hua Wang$^{1}$}
\email{wangxueh@mail.sysu.edu.cn}
\address{$^1$State Key Laboratory of Optoelectronic Materials and Technologies, Sun Yat-sen
University, Guangzhou 510275, China}
\address{$^2$College of Physics Science and Information Engineering, Jishou University,
Jishou 416000, China}
\address{$^3$Beijing National Laboratory for Condensed Matter Physics,
Institute of Physics, Chinese Academy of Sciences, Beijing 100190
China}
\date{\today}

\begin{abstract}
Dipole-dipole interaction between two two-level `atoms' in photonic crystal
nanocavity is investigated based on finite-difference time domain algorithm.
This method includes both real and virtual photon effects and can be applied
for dipoles with different transition frequencies in both weak and strong
coupling regimes. Numerical validations have been made for dipoles in vacuum
and in an ideal planar microcavity. For dipoles located in photonic crystal
nanocavity, it is found that the cooperative decay parameters and the
dipole-dipole interaction potential strongly depend on the following four
factors: the atomic position, the atomic transition frequency, the resonance
frequency, and the cavity quality factor. Properly arranging the positions
of the two atoms, we can acquire equal value of the cooperative decay
parameters and the local coupling strength. Large cooperative decay
parameters can be achieved when transition frequency is equal to the
resonance frequency. For transition frequency varying in a domain of the
cavity linewidth around the resonance frequency, dipole-dipole interaction
potential changes continuously from attractive to repulsive case. Larger
value and sharper change of cooperative parameters and dipole-dipole
interaction can be obtained for higher quality factor. Our results provide some manipulative approaches for dipole-dipole interaction with potential application in various fields such as quantum computation and quantum information processing based on solid state nanocavity and quantum dot system.
\end{abstract}

\maketitle

\section{I. INTRODUCTION}

Since Purcell predicted spontaneous emission rate could be changed by
electromagnetic environment in 1946 \cite{purcell}, the effect of
electromagnetic field on radiation properties has been thoroughly
investigated \cite%
{Barton,AgarwalA121475,Kleppner,John,Yokoyama,Knoll,Wang,Yoshie,Thon}, and
classified in the category of cavity quantum electrodynamics (QED).
Characteristics of QED have been greatly investigated both theoretically and
experimentally \cite{Khitrova,Birnbaum,Greentree,Goy,Walther,Andersen}, and
many kinds of devices \cite{Braun,Panajotov} based on this theory have been
developed. Concepts such as enhanced and inhibited spontaneous emission \cite%
{Kleppner,John}, reversible spontaneous emission \cite{Wang}, photon
blockade \cite{Birnbaum}, micromasers \cite{Meschede}, low threshold lasers
\cite{Painter,Noda}, etc., have become very familiar.

Dipole-dipole interaction could also be greatly modulated by the
electromagnetic environment. Photon emitted by one dipole could be absorbed
by the other or vice versa. The strength of interaction is decided by photon
emission, transmission and absorption. Many different kinds of
electromagnetic environment can be used to control or change these
characteristics, such as vacuum \cite%
{AgarwalA121475,LehmbergA2883,Agarwal,Hettich,Puhan}, optical cavities \cite%
{Goldstein,Kobayashi,LehmbergA2889,GSagarwal}, optical lens \cite{Rist},
dielectric droplet \cite{Arnold}, photonic materials \cite%
{Bay,Shughesdd,Gallardo,Laucht,Reitzenstein,Weiler,Kim}, metal surface \cite%
{metasurface1,metasurface2,metasurface3,metasurface4,metasurface5},
metamaterial \cite{metamaterial,metamaterial1} and so on. For example,
optical lens and waveguide have been designed to collect the emission photon
and transfer it to the other dipole. Optical cavity or metal surface can
enhance the emission or absorption rate roughly by the ratio of quality
factor $Q$ and the mode voluum $V$.

Photonic crystal nanocavity is one of the promising platform to investigate
dipole-dipole interaction, because local coupling strength of dipole and
photon can be tailored and integrated to photonic crystal waveguide is
extremely convenient. High quality factor $Q=2.5\times 10^{6}$ and small
mode volume $V\sim (\lambda /n)^{3}$ have been realized for photonic crystal
cavity \cite{SNodaQ}. Numerical investigations show that ultra-high quality
factor with $Q\sim 10^{9}$ can be designed through finely tuning the
scatters around the cavity with little change of the mode volume \cite%
{SnodaQT}. Furthermore, static and ultra fast dynamic control of the
cavity frequency and the quality factor have been achieved \cite%
{Vignolini,Mosor,SnodaSQ,SnodaDQ,TanabeDF}. On the other hand, temperature
\cite{Yoshie}, strainin \cite{strainin}, electric field \cite{electric field}%
, magnetic field \cite{magnetic field,magnetic field1} are much suitable for
fine tuning the energe levels of dipoles located at certain position in
solid system.

Recent studies show that this kind of interaction could be used to implement
quantum entanglement preparation and quantum information processing \cite%
{Lukin0,Lukin1,Lukin2,Isenhower,ZhengSB,ShughesL05,ShughesOE,Gonzalez-Tudela}%
, cooperative radiation \cite{LehmbergA2889,Ruostekoski}, F\"{o}rster energe
transfer \cite{Xie,Youl}, dipole nanolaser \cite{Protsenko}, and so on.
Furthermore, some novel quantum phenomena\cite%
{Reinhard,Johnson,Saquet,Harlander,Miranda} has been found. All of these
applications and phenomena are related to the cooperative decay parameters
or dipole-dipole interaction potential. According to Eq. (\ref{6c}),
dipole-dipole interaction potential can be got through the cooperative decay
parameters.

In the previous theoretical studies, mode-expansion method \cite%
{Goldstein,Kobayashi,Agarwal,Bay,Rist,Arnold} or Green function method \cite%
{Schmid,GSagarwal,ShughesL05,ShughesOE,Gonzalez-Tudela} is often adopted to
investigate this cooperative decay parameters. These two methods work well
for electromagnetic environment with perfect boundary condition. Because of
the extremely complexity of finding eigen-mode, mode-expansion method can be
used only for simple case such as vacuum or planar cavity. Besides, the
exact analytic Green function is also hard to be obtained for complex
electromagnetic environments. Numerical method is necessary for studying
this kind of interaction in photonic crystal nanostructure.

In this paper, we put forward a simple numerical method to investigate
dipole-dipole interaction in photonic crystal nanocavity through
finite-difference time domain algorithm (FDTD). We calculate the collective
and individual radiation rates of classical dipoles by directly solving
Maxwell's equations in real space with a free-space boundary condition. By
using the result of two dipoles radiation rate minus the sum of the two
individual radiation rates, we get the cooperative decay parameters and
dipole-dipole interaction potential. A similar method has been used for
local density of photonic states calculation \cite{Dowling,XuY,Hwang}. It is
direct for structures with arbitrary shape and physical quantities such as
radiation power at different frequencies can be got in a single simulation
run. Numerical results in vacuum and planar conductor cavity show that our
method works well for this kind of investigation. For dipoles located in
photonic crystal cavity, the strength of dipole-dipole interaction depends
heavily on the following four factors: the atomic position, the transition
frequency, quality factor, mode volume and the cavity frequency. Effect of
these factors on the interaction strength are also shown.

This paper is organized as follows: In Sec. II, we give the model of
dipole-dipole interaction with some illuminations. Numerical method is
presented in Sec. III. Validation of this method is presented in Sec. IVA
for dipoles in vacuum and in planar microcavity. In Sec. IVB, applying our
method to photonic crystal nanocavity, we investigate the effect of the
atomic position, the transition frequency, quality factor and the cavity
frequency on the dipole-dipole interaction strength. In Sec. V, we give a
brief conclusion.

\section{II. Model}

\begin{figure}[htbp]
\includegraphics[scale=0.72]{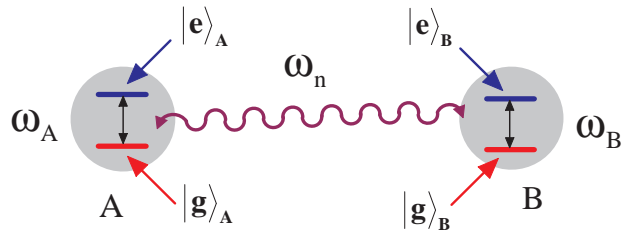}
\caption{(Color online). Schematic diagram of dipole-dipole interaction.
Consider two two-level `atoms' A and B. Atom A (B) has two states, the
ground state $|g_{A}\rangle $ ($|g_{B}\rangle $) and the excited state $%
|e_{A}\rangle $ ($|e_{B}\rangle $), with a transition frequency $\protect%
\omega _{A}$ ($\protect\omega _{B}$). Atom A is in the excited state $%
|e_{A}\rangle $ and atom B is in the ground state $|g_{B}\rangle $. They
both interact with electromagnetic field with eigen frequency $\protect%
\omega _{n}$. Analogous to the process of Lamb shift where both virtral and
real photon exchange between the atom and the quantum electromagnetic field
take effect, photom exchange between the two atoms contribute to
dipole-dipole interaction. }
\end{figure}

Schematic diagram of dipole-dipole interaction is illustrated in Fig. 1.
There are two two-level `atoms' A and B located at $\mathbf{r}_{A}$ and $%
\mathbf{r}_{B}$ respectively and they both interact with the electromagnetic
field with eigen frequency $\omega _{n}$. The Atom A (B) has two states, the
ground state $|g_{A}\rangle $ ($|g_{B}\rangle $) and the excited state $%
|e_{A}\rangle $ ($|e_{B}\rangle $), with a transition frequency $\omega _{A}$
($\omega _{B}$). The Hamiltonian of the system in the rotation wave approach
reads \cite{Louisell}:

\begin{align}
H& =H_{0}+V,  \notag \\
H_{0}& =\hbar \sum_{i=A,B}\omega _{i}|e_{i}\rangle \langle e_{i}|+\hbar
\sum_{n}\omega _{n}a_{n}^{\dagger }a_{n},  \notag \\
& V=\hbar \sum_{i=A,B}\sum_{n}[g_{n}(\mathbf{r}_{i})a_{n}^{\dagger
}|g_{i}\rangle \langle e_{i}|+c.c].  \label{1}
\end{align}

\noindent where $a_{n}^{\dagger }$ and $a_{n}$ are, the photonic creation
and annihilation operators, $\omega _{n}$ is the frequency of the eigen mode
of the electromagnetic field, $g_{n}(\mathbf{r}_{i})=i\omega
_{j}(2\varepsilon _{0}\hbar \omega _{n})^{-1/2}\mathbf{E}_{n}(\mathbf{r}%
_{i}).\mathbf{u}_{i}$ ( $i=A,B$ ), is the the coupling coefficient, $\mathbf{%
E}_{n}(\mathbf{r})$ is the eigen mode, $\mathbf{u}_{i}$ is the transition
dipole moments of atoms $i$, and $\varepsilon _{0}$ is the vacuum
permittivity. In Eq. (\ref{1}), $H_{0}$ is the noninteracting Hamiltonian,
and $V$ represents the interaction between dipole and electromagnetic field.
There are three states for the system considered: (1) atom A is in the
excited state, and atom B is in the ground state, without any photon, i.e., $%
|a\rangle =|e_{A},g_{B},0\rangle $, (2) atom B is in the excited state, and
atom A is in the ground state, without any photon, i.e., $|b\rangle
=|g_{A},e_{B},0\rangle $, (3) both atom A and atom B are in the ground
state, with a photon of frequency $\omega _{n}$, i.e., $|c_{n}\rangle
=|g_{A},g_{B},1_{n}\rangle $.

The initial state is prepared in $|a\rangle$. Then, the state of the system
evolves as

\begin{equation}
|\psi (t)\rangle =a(t)|a\rangle +b(t)|b\rangle
+\sum_{n}c_{n}(t)|c_{n}\rangle \equiv U(t)|a\rangle .  \label{0}
\end{equation}

\noindent where $U(t)$ is the evolution operator and $a(t)=\langle
a|U(t)|a\rangle $, $b(t)=\langle b|U(t)|a\rangle $, $c_{n}(t)=\langle
c_{n}|U(t)|a\rangle $. To derive the equation for the atom-field dynamics
nonperturbatively, the resolvent operator $G(z)=1/(z-H)$ is adopted \cite%
{Cohen-Tannoudji}. The corresponding advanced and retarded propagators are $%
G^{\pm }(E)=\lim_{\eta \rightarrow 0^{+}}G(E\pm i\eta )$. Then the evolution
operator can be expressed by

\begin{equation}
U(t)=\int_{-\infty }^{+\infty }d\omega \lbrack G^{-}(\omega )-G^{+}(\omega
)]\exp (-i\omega t)/2\pi i.  \label{02}
\end{equation}

The dynamic properties of this system are governed by $G_{aa}^{\pm }(\omega
) $ and $G_{ba}^{\pm }$. The matrix elements of the resolvent reads

\begin{align}
(z-\hbar \omega _{A})G_{aa}& =1+\sum_{n}V_{ac}G_{ca},  \notag \\
(z-\hbar \omega _{B})G_{ba}& =\sum_{n}V_{bc}G_{ca},  \notag \\
(z-\hbar \omega _{n})G_{ca}& =V_{ca}G_{aa}+V_{cb}G_{ba}.  \label{2}
\end{align}

\noindent where $G_{aa}=\langle a|G(z)|a\rangle$, $G_{BA}=\langle
b|G(z)|a\rangle$, $G_{ca}=\langle c_{nk}|G(z)|a\rangle$ and $%
V_{ac}=V_{ca}^{\ast}=\langle a|V|c\rangle$, $V_{bc}=V_{cb}^{\ast}=\langle
b|V|c\rangle$

Eliminating $G_{ca}$, we have

\begin{align}
G_{aa}(z)& =(z-\hbar \omega _{B}-W_{BB}(z))/\Xi ,  \notag \\
G_{ba}(z)& =W_{BA}(z)/\Xi .  \label{3}
\end{align}
$\Xi $ is given by

\begin{equation}
\Xi =[z-\hbar \omega _{A}-W_{AA}][z-\hbar \omega _{B}-W_{BB}]-W_{AB}W_{BA}
\label{4}
\end{equation}

The local coupling between atom and the field ($W_{AA},W_{BB}$ ) or the
dipole-dipole coupling between atom A and atom B ($W_{AB},W_{BA}$) are
denoted by
\begin{equation}
W_{ij}(z,\mathbf{r}_{i},\mathbf{r}_{j})=\hbar \sum_{n}\frac{g_{n}^{\ast }(%
\mathbf{r}_{i})g_{n}(\mathbf{r}_{j})}{z-\hbar \omega _{n}}.  \label{5}
\end{equation}
Clearly, these terms can be written as
\begin{subequations}
\begin{align}
W_{ij}^{\pm }(\hbar \omega ,\mathbf{r}_{i},\mathbf{r}_{j})& =\hbar \lbrack
\Delta _{ij}(\omega ,\mathbf{r}_{i},\mathbf{r}_{j})\mp i\frac{\Gamma
_{ij}(\omega ,\mathbf{r}_{i},\mathbf{r}_{j})}{2}],\text{ } \\
\Gamma _{ij}(\omega ,\mathbf{r}_{i},\mathbf{r}_{j})& =2\pi
\sum_{n}g_{n}^{\ast }(\mathbf{r}_{i})g_{n}(\mathbf{r}_{j})\delta (\omega
-\omega _{n}),  \label{6} \\
\Delta _{ij}(\omega ,\mathbf{r}_{i},\mathbf{r}_{j})& =\frac{1}{2\pi }\wp
\int_{0}^{+\infty }dz\frac{\Gamma _{ij}(z,\mathbf{r}_{i},\mathbf{r}_{j})}{%
\omega -z}.  \label{6c}
\end{align}

To better understand the physics underlying of these equations, we make some
illumination. We are able to self-consistently determine the roots of $\Xi
=0 $ (Eq. (\ref{4})). The real parts are corresponding to the energy levels
of these two atoms and the imaginary parts are the lifetime. Different from
the usual Markov approximation method where $W_{ij}$ $(i,j=A,B)$ are
independent of $\omega $ and have been replaced by their approximate value $%
W_{ii}(\omega _{i})$ and $W_{ij}((\omega _{i}+\omega _{j})/2)$, this method
is non-Markov and this self-consistent process can give the rigorous
dipole-dipole coupling. We can also make use of the Markov approximation\
values $W_{ij}$ $(i,j=A,B)$ in Eq. (\ref{4}) and get the same results as
usual. However, for complex electromagnetic environment such as photonic
crystal or photonic crystal nanocavity, $W_{ij}$ may vary sharply around
some frequency and non-Markov is necessary. Owing to $W_{ij}$, equation. (%
\ref{4}) also shows that both the energy lever and the lifetime of the
states are spliting. Different coupling $W_{ij}$ makes different energy
lever spliting and different lifetime spliting. Dipole blockade,\ which have
been widely investigated in quantum information processing needs large
energy gap. Efficient superradiant emission, steady state entanglement
preparation and fast F\"{o}rster energy transfer needs large lifetime
spliting. Inspired by these novel application, accurate values for $W_{ij}$
are expected. Furthermore, equation (\ref{5}) shows that all of the photonic
eigen modes contribute to the local coupling and dipole-dipole interaction
and both real and virtual photon effects have been taken into account.

Insert $g_{n}(\mathbf{r}_{i})=i\omega _{i}(2\varepsilon _{0}\hbar \omega
_{n})^{-1/2}\mathbf{E}_{n}(\mathbf{r}_{i}).\mathbf{u}_{i}$ $(i=A,B)$ into
Eq. (\ref{6}) for $i\neq j$, and define $s_{i,j}(\omega )\equiv \pi \omega
_{i}\omega _{j}u_{i}u_{j}/(\varepsilon _{0}\hbar \omega )$ $(i=A,B)$, where $%
u_{i}$ and $\widehat{\mathbf{u}}_{i}$ are the size and the unit vector of
transition dipole $\mathbf{u}_{i}$. Then, the cooperative decay parameters
reads:
\end{subequations}
\begin{equation}
\Gamma _{ij}(\omega ,\mathbf{r}_{i},\mathbf{r}_{j})=s_{i,j}(\omega )\sum_{n}%
\mathbf{E}_{n}^{\ast }(\mathbf{r}_{i}).\widehat{\mathbf{u}}_{i}\mathbf{E}%
_{n}(\mathbf{r}_{j}).\widehat{\mathbf{u}}_{j}\delta (\omega -\omega _{n}).
\label{7}
\end{equation}%
For simplicty, we denote $\Gamma _{ij}(\omega )$ and $\Delta _{ij}(\omega )$
for $\Gamma _{ij}(\omega ,\mathbf{r}_{i},\mathbf{r}_{j})$ and $\Delta
_{ij}(\omega ,\mathbf{r}_{i},\mathbf{r}_{j})$ respectively.

From equations 2.12a to 2.14a of reference \cite{Glauber}, it is easy to see
that if $\left\{ \mathbf{E}_{n}(\mathbf{r})\right\} $ compose a complete set
of eigen mode, $\left\{ \mathbf{E}_{n}^{\ast }(\mathbf{r})\right\} $ is also
a complete set of eigen mode. Then
\begin{equation}
\Gamma _{ij}(\omega )=s_{i,j}(\omega )\sum_{n}\mathbf{E}_{n}^{\ast }(\mathbf{%
r}_{i}).\widehat{\mathbf{u}}_{i}\mathbf{E}_{n}(\mathbf{r}_{j}).\widehat{%
\mathbf{u}}_{j}\delta (\omega -\omega _{n}).
\end{equation}%
could also be written as:
\begin{equation}
\Gamma _{ij}(\omega )=s_{i,j}(\omega )\sum_{n}\mathbf{E}_{n}(\mathbf{r}_{i}).%
\widehat{\mathbf{u}}_{i}\mathbf{E}_{n}^{\ast }(\mathbf{r}_{j}).\widehat{%
\mathbf{u}}_{j}\delta (\omega -\omega _{n}).
\end{equation}%
So we have :

\begin{equation}
\Gamma _{ij}(\omega )=\frac{s_{i,j}(\omega )}{2}\sum_{n}[\mathbf{E}_{n}(%
\mathbf{r}_{i}).\widehat{\mathbf{u}}_{i}\mathbf{E}_{n}^{\ast }(\mathbf{r}%
_{j}).\widehat{\mathbf{u}}_{j}+h.c.]\delta (\omega -\omega _{n}).  \label{8}
\end{equation}

Once we get the cooperative decay parameters $\Gamma_{ij}(\omega)$, the
dipole-dipole interaction potential $\Delta_{ij}(\omega)$ can be achieved
through Eq. (\ref{6c}).

\section{III. METHOD}

Here we propose a new method to rigorously get the cooperative decay
parameters based on finite-difference time domain algorithm. We show that
dipole-dipole interaction can be got through calculating the collective and
individual radiation rates of classical dipoles. We begin with Maxwell
equations:
\begin{align}
\nabla \times \mathbf{E}(\mathbf{r},t)& =-\frac{\partial \mathbf{B}(\mathbf{r%
},t)}{\partial t},  \notag \\
\nabla \times \mathbf{B}(\mathbf{r},t)& =\mu _{0}\varepsilon (\mathbf{r})%
\frac{\partial \mathbf{E}(\mathbf{r},t)}{\partial t}+\mu _{0}\frac{\partial
\mathbf{P}(\mathbf{r},t)}{\partial t},  \notag \\
\nabla \cdot \varepsilon (\mathbf{r})\mathbf{E}(\mathbf{r},t)& =\rho (%
\mathbf{r},t),  \notag \\
\nabla \cdot \mathbf{B}(\mathbf{r},t)& =0.
\end{align}%
Expand $\mathbf{E}(\mathbf{r},t)=\sum_{n}\alpha _{n}(t)\mathbf{E}_{n}(%
\mathbf{r})$ where $\mathbf{E}_{n}(\mathbf{r})$ is the same as the eigen
modes at the quantum analysis section. If we let $\mathbf{P}(\mathbf{r}%
,t)=e^{-i\omega _{0}t}\mathbf{u}(\mathbf{r})$, then $\alpha _{n}(t)$
satisfies the following equation:
\begin{equation}
\ddot{\alpha _{n}(t)}+\omega _{n}^{2}\alpha _{n}(t)=-\omega
_{0}^{2}e^{-i\omega _{0}t}\int d\mathbf{ru}(\mathbf{r})\cdot \mathbf{E}%
_{n}^{\ast }(\mathbf{r}).  \label{12}
\end{equation}%
Through Eq. (\ref{12}), we get:
\begin{equation}
\alpha _{n}(t)=\lim_{\eta \rightarrow 0^{+}}\frac{-\omega
_{0}^{2}exp(-i\omega _{0}t)}{\omega _{n}^{2}-\omega _{0}^{2}+i\eta }\int d%
\mathbf{ru}(\mathbf{r})\cdot \mathbf{E}_{n}^{\ast }(\mathbf{r}).
\end{equation}

If $\mathbf{u}(\mathbf{r})=\sum\limits_{i}\widehat{\mathbf{u}}_{i}\delta (%
\mathbf{r}-\mathbf{r}_{i})$, the emission power is given by :

\begin{equation}
P(\omega _{0})=\frac{\pi }{4}\omega _{0}^{2}\sum\limits_{n}\mid
\sum\limits_{i}\widehat{\mathbf{u}}_{i}\cdot \mathbf{E}_{n}^{\ast }(\mathbf{r%
}_{i})\mid ^{2}\delta (\omega _{0}-\omega _{n}).  \label{16}
\end{equation}

In the two dipoles instance, i.e., $i=A$, $B$, the total power is%
\begin{equation}
P^{AB}(\omega _{0})=\frac{\pi }{4}\omega _{0}^{2}\sum\limits_{n}\mid
\sum\limits_{i=A,B}\widehat{\mathbf{u}}_{i}\cdot \mathbf{E}_{n}^{\ast }(%
\mathbf{r}_{i})\mid ^{2}\delta (\omega _{0}-\omega _{n}).  \label{17}
\end{equation}

If there is only one dipole, i.e., $i=A$ or $i=B$, then

\begin{align}
P^{A}(\omega _{0})& =\frac{\pi }{4}\omega _{0}^{2}\sum\limits_{n}\mid
\widehat{\mathbf{u}}_{A}\cdot \mathbf{E}_{n}^{\ast }(\mathbf{r}_{A})\mid
^{2}\delta (\omega _{0}-\omega _{n}),  \label{18} \\
P^{B}(\omega _{0})& =\frac{\pi }{4}\omega _{0}^{2}\sum\limits_{n}\mid
\widehat{\mathbf{u}}_{B}\cdot \mathbf{E}_{n}^{\ast }(\mathbf{r}_{B})\mid
^{2}\delta (\omega _{0}-\omega _{n}).  \label{19}
\end{align}

Combining Eq. (\ref{17}), Eq. (\ref{18}) and Eq. (\ref{19}), we difine the
cooperative emission power as:

\begin{align}
P_{co}(\omega _{0})& \equiv P^{AB}(\omega _{0})-P^{A}(\omega
_{0})-P^{B}(\omega _{0}),  \notag \\
& =\frac{\pi \omega _{0}^{2}}{4}\sum\limits_{n}[\widehat{\mathbf{u}}%
_{A}\cdot \mathbf{E}_{n}^{\ast }(\mathbf{r}_{A})\widehat{\mathbf{u}}%
_{B}\cdot \mathbf{E}_{n}(\mathbf{r}_{B})+  \notag \\
& h.c.]\delta (\omega _{0}-\omega _{n}).  \label{20}
\end{align}

Through Eq. (\ref{20}) and Eq. (\ref{8}), we found that:%
\begin{equation}
\frac{\Gamma _{ij}^{c}(\omega )}{\Gamma _{ii}^{0}(\omega )}=\frac{\mu
_{i}\mu _{j}\omega _{i}\omega _{j}}{\mu _{i}^{2}\omega _{i}^{2}}\frac{%
P_{co}^{c}(\omega )}{2P_{0}^{v}(\omega )}.  \label{21}
\end{equation}%
where $\Gamma _{ij}^{c}(\omega )$ and $P_{co}^{c}(\omega )$ correspond to
cooperative decay parameters and cooperative emission power of two unit
classical dipoles in complex environment. $P_{0}^{v}(\omega )$ correspond to
the emission power of a unit classical dipole in vacuum. $\Gamma
_{ii}^{0}(\omega )$ is the local coupling strength for an two level atom
with transition dipole moment $\mathbf{u}_{i}$ and transition frequency of
bare atom $\omega _{i}$ in vacuum. Using the result of \cite{Wang}, we get $%
\Gamma _{ii}^{0}(\omega )/(u_{i}^{2}\omega _{i}^{2})=\omega /(3\pi
\varepsilon _{0}\hbar c^{3})$.

In order to make the cooperative decay parameters more clearly, we define

\begin{eqnarray}
\eta (\omega ) &\equiv &\frac{P_{co}^{c}(\omega )}{2P_{0}^{v}(\omega )},
\notag \\
\alpha _{i} &\equiv &\frac{u_{i}^{2}\omega _{i}^{2}}{3\pi \varepsilon
_{0}\hbar c^{3}}.
\end{eqnarray}%
where $\alpha _{i}$ is totally decided by the two level atom i. Then we get

\begin{equation}
\Gamma _{ij}^{c}(\omega )=\sqrt{\alpha _{i}}\sqrt{\alpha _{j}}\eta (\omega
)\omega .  \label{22}
\end{equation}

Equation. (\ref{22}) is the main content of our method. $\eta (\omega )$ can
be obtained through calculating $P_{co}^{c}(\omega )$ and $P_{0}^{v}(\omega )
$ by FDTD method. Utilizing Eq. (\ref{6c}), we can get the dipole-dipole
interaction potential $\Delta _{ij}(\omega )$. Furthermore, we see that our
method may be generalized to study many dipole interactions through this
similar procedure.

\section{IV. NUMERICAL RESULTS AND DISCUSSION}

\subsection{A. Cooperative decay parameters in vacuum and planar microcavity}

\begin{figure}[tbph]
\includegraphics[scale=0.5]{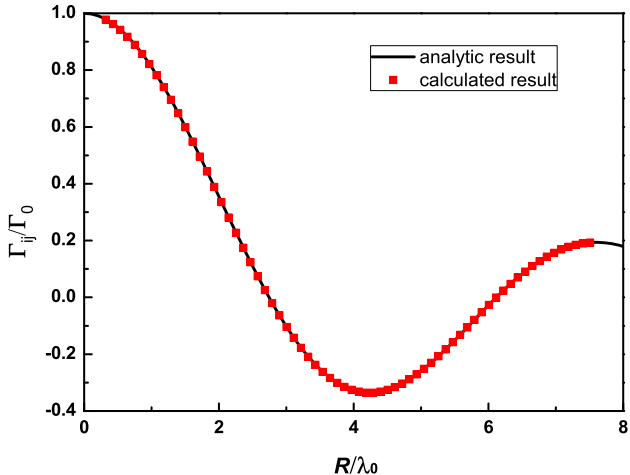}
\caption{(Color online). Comparison between our numerical results and the
analytic results of cooperative decay parameters for dipoles located in
vacuum. Cooperative decay parameters $\Gamma _{ij}/\Gamma _{0}$ as a
function of the interatomic separation $R/\protect\lambda _{0}$ when $%
\protect\omega =\protect\omega _{0}=2\protect\pi c/\protect\lambda _{0}$;
Black line is for the analytic results (Eq. (\protect\ref{23})) and red
boxed dots represent the calculated results (Eq. (\protect\ref{22})). The
orientation of the two transition dipole moments are the same and
perpendicular to their separation.}
\label{ssp2}
\end{figure}

Before applying our method to investigate dipole-dipole interaction in
photonic crystal nanocavity, we first numerically validate our approach in
the case for dipoles in vacuum or planar microcavity where analytic
expression for the cooperative decay parameters $\Gamma _{ij}^{c}(\omega )$
can be easily got through mode-expansion method. In this section, for
simplicity, we set $\omega _{A}=\omega _{B}=\omega _{0}$ and $\alpha
_{A}=\alpha _{B}=\alpha _{0}=\omega _{0}^{2}u_{i}^{2}/3\pi \varepsilon
_{0}\hbar c^{3}$. In vacuum, the two transition dipole moments are parallel
to the x axis and the interatomic separation axis is along the z axis. The
eigen-function of vacuum reads $\mathbf{E}_{\mathbf{k},l}(\mathbf{r})=\hat{e}%
_{\mathbf{k}}^{l}e^{i\mathbf{k}\cdot \mathbf{r}}$. Inserting this into Eq. (%
\ref{8}) and making use of $\sum_{l=1,2}\hat{e}_{\mathbf{k},i}^{l}\hat{e}_{%
\mathbf{k},j}^{l}=\delta _{ij}-\mathbf{\hat{k}}_{i}\mathbf{\hat{k}}_{j}$, we
get the analytic cooperative decay parameters $\Gamma _{ij}(\omega )$:
\begin{equation}
\Gamma _{ij}(\omega )=\frac{3\omega \Gamma _{0}(\omega _{0})}{2\omega _{0}}[%
\frac{\sin (\omega R/c)}{(\omega R/c)}+\frac{\cos (\omega R/c)}{(\omega
R/c)^{2}}-\frac{\sin (\omega R/c)}{(\omega R/c)^{3}}].  \label{23}
\end{equation}%
where $R=\left\vert \mathbf{r}_{i}-\mathbf{r}_{j}\right\vert $ is the
distance between the two dipoles. If $R=0$, for the two dipole located at
the same place, $\Gamma _{ij}(\omega _{0})=\Gamma _{0}(\omega _{0})=\omega
_{0}^{3}u_{i}^{2}/3\pi \varepsilon _{0}\hbar c^{3}$ which is the radiation
rate of the dipole in vacuum. We set $\omega =\omega _{0}=2\pi c/\lambda $
and increase the interatomic separation $R$ to simplify our discussion. In
Fig. \ref{ssp2}, black line is for the analytic results (Eq. (\ref{23})) and
red boxed dots represent the calculated results (Eq. (\ref{22})). It can be
clearly seen that $\Gamma _{ij}(\omega )$ got through our method agrees very
well with the analytic results.

\begin{figure}[tbph]
\includegraphics[scale=0.8]{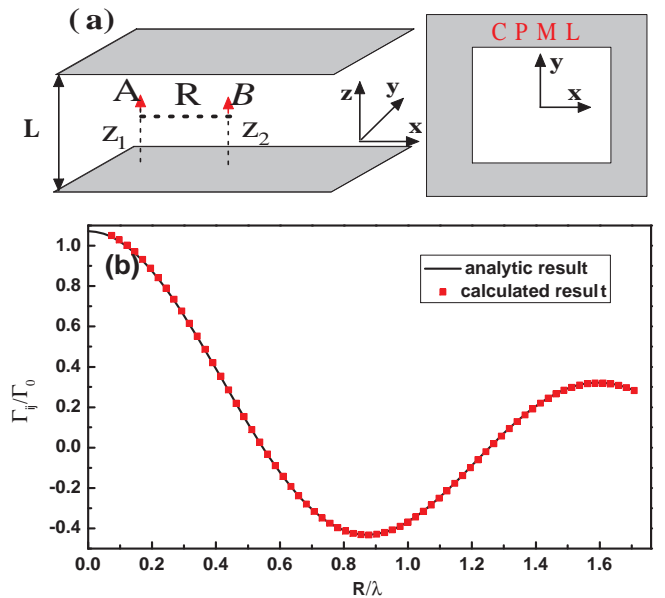}
\caption{(Color online). Sketch of setup and the comparison between our
numerical results and the analytic results of cooperative decay parameters
for dipoles located in vacuum. (a) Schematic diagram of the planar cavity
with a top view of the computation domain at the right side. The two dipoles
locate at the center. The orientation of the two transition dipole moments
are the same and along z (red arrowhead). They are arranged along x axis and
their separation is $R$. (b) Cooperative decay parameters $\Gamma
_{ij}/\Gamma _{0}$ as a function of the interatomic separation $R$ when $%
z_{1}=z_{2}=L/2$, $L/\protect\lambda _{0}=L/\protect\lambda =0.7$ ; black
line is for the analytic results from Eq. (\protect\ref{24}) and red boxed
dots represent the calculated from Eq. (\protect\ref{22}).}
\label{ssp3}
\end{figure}
Similar result is also got when the two point dipoles are located in a
planar nanocavity. Here, the cooperative decay parameters $\Gamma
_{ij}(\omega )$ could also be got analytically. The structure is shown in
Fig. \ref{ssp3}(a). The origin is at the center for x, y axis and at the
bottom conductor for z axis. The CPML encloses the domain, as can be seen
from the top view at the right-hand side. The orientation of the two
transition dipole moments are the same and along z (red arrowhead). They are
arranged along x axis and their separation is $R$. The analytic cooperative
decay parameter $\Gamma _{ij}(\omega )$ for these two dipoles is:%
\begin{align}
\Gamma _{ij}(\omega )& =\Gamma _{0}(\omega _{0})(\frac{3\lambda _{0}}{8\pi L}%
)\{\int_{0}^{2\pi }d\theta \cos (\frac{2\pi R}{\lambda }\cos \theta )+
\notag \\
& \sum\limits_{n=1}^{[2L/\lambda ]}2[1-(\frac{n\lambda }{2L})^{2}]\cos (%
\frac{n\pi z_{1}}{L})\cos (\frac{n\pi z_{2}}{L})\}\times  \notag \\
& \int_{0}^{2\pi }d\theta \cos (2\pi R\sqrt{(\frac{1}{\lambda })^{2}-(\frac{n%
}{2L})^{2}}\cos \theta ).  \label{24}
\end{align}

\noindent where $\omega =2\pi c/\lambda $, $[2L/\lambda ]$ is the largest
integer less than $2L/\lambda $, $\Gamma _{0}(\omega _{0})$ is spontaneous
emission rate in vacuum for any one of the two same dipoles, and $z_{i}$ is
the coordinate of z for dipole i, here, we set $z_{1}=z_{2}=L/2$, $R$ is the
distance between the two dipoles in xy plane. For $R=0$, this leads to the
well-known results

\begin{equation}
\Gamma_{z}(\omega_{0})=\Gamma_{0}(\omega_{0})(\frac{3\lambda_{0}}{4L}%
)\{1+\sum\limits_{n=1}^{[2L/\lambda_{0}]}2[1-(\frac{n\lambda_{0}}{2L}%
)^{2}]cos^{2}(\frac{n\pi}{2})\}.
\end{equation}

In the calculation, we set $L/\lambda _{0}=L/\lambda =0.7$, and $R/L$ is
gradually increased from 0 to 49/41 for simplicity. Cooperative decay
parameters $\Gamma _{ij}/\Gamma _{0}$ as a function of the interatomic
separation $R$ is displayed in Fig. \ref{ssp3}(b). black line is for the
analytic results from Eq. (\ref{24}) and red boxed dots represent the
calculated from Eq. (\ref{22}). We find that the numerical result from our
method also agrees very well with the analytic solution.

\begin{figure}[tbph]
\includegraphics[scale=0.6]{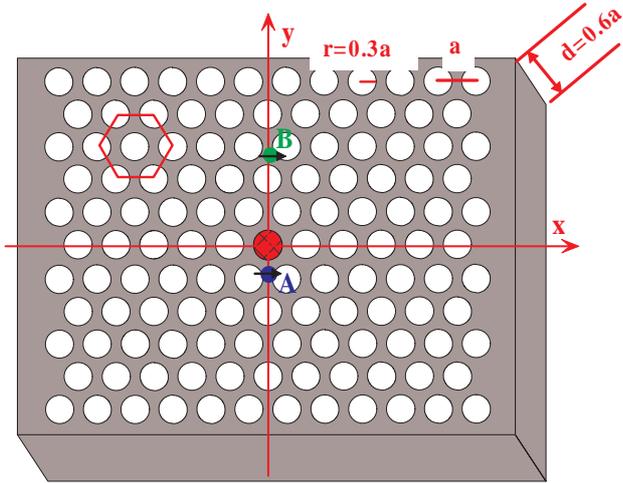}
\caption{(Color online). The sketch of the photonic crystal nanocavity. It
consists of a thin dielectric slab with air holes arranged as triangular
lattice. The lattice constant is $a$, the radii of the air hole is $r=0.3a$
and the slab height is $d=0.6a$. The refractive index of the slab is 3.4.
There is a defect hole in the center with refractive index $n_{def}=2.4$.
Its radii is the same as the air hole. The origin of the axes is set at the
center of the photonic crystal nanocavity. The two dipoles are located at
point A and B on the center plane of the slab and the two dipole moments are
parallel to the x axis.}
\label{ssp4}
\end{figure}

\subsection{B. Dipole-dipole interaction in photonic crystal nanocavity}

Through the above two numerical tests, we clearly shown that our method
could be able to efficiently and exactly deal with the dipole-dipole
interaction. In the following of this section, dipole-dipole interaction in
photonic crystal cavity (PCs) has been numerically studied. The sketch of
the photonic crystal cavity is displayed in Fig. \ref{ssp4}. It consists of
a thin dielectric slab with air holes arranged as triangular lattice. The
lattice constant is $a$, the radii of the air hole is $r=0.3a$ and the slab
height is $d=0.6a$. The refractive index of the slab is 3.4. There is a
defect hole in the center with refractive index $n_{def}=2.4$. Its radii is
the same as the air hole. The two dipoles are located at point A and B on
the center plane of the slab and the two dipole moments are parallel to the
x axis. The origin of the axes is set at the center of the PCs cavity. For
convenience, the special points are also drawn in Fig. \ref{ssp4}. `Atom' A
located at $(0,-11/15,0)a$, `Atom' B located $(0,R-11a/15,0)$, where R
represents the separation and varies from $a/15$ to $50a/15$. The two
transition dipole moments are parallel to the x axis. Thanks to the scaling
law, the cooperative decay parameters $\Gamma _{ij}(\omega )$, dipole-dipole
interaction potential $\Delta _{ij}(\omega )$ and frequency $\omega $ in
unit of $2\pi c/a$ are dimensionless.

\begin{figure}[htbp]
\includegraphics[scale=0.465]{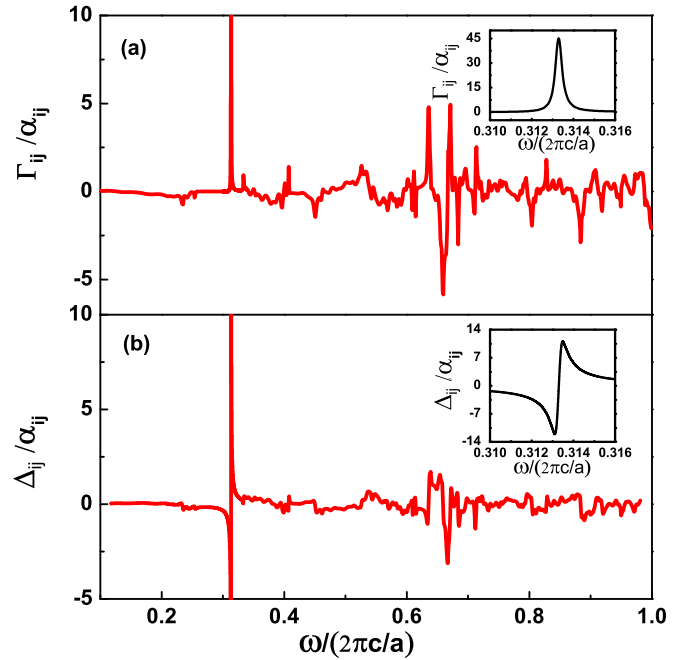}
\caption{(Color online). The frequency dependent characters for cooperative
decay parameters and dipole-dipole interaction potential. (a) The
cooperative decay parameters $\Gamma _{ij}(\protect\omega )$ and (b) the
dipole-dipole interaction potential $\Delta _{ij}(\protect\omega )$ versus $%
\protect\omega $ for atom A located at $(0,-11/15,0)a$, atom B located at $%
(0,11/15,0)a$ in the photonic crystal nanocavity. $\Gamma _{ij}(\protect%
\omega )$ and $\Delta _{ij}(\protect\omega )$ are in unit of $\protect\alpha %
_{ij}$ ($\protect\alpha _{ij}\equiv \protect\sqrt{\protect\alpha _{i}}%
\protect\sqrt{\protect\alpha _{j}}2\protect\pi c/a$). The $\protect\omega $
is in unit of $2\protect\pi c/a$. The two dipole moments are parallel and
along the x axis. The insets show the behavior around the defect frequency ($%
\protect\omega _{c}=0.3133(2\protect\pi c/a)$) of the cavity. The two
transition dipoles are parallel and along the x axis.}
\label{ssp5}
\end{figure}

The frequency dependent characters for cooperative decay parameters $\Gamma
_{ij}(\omega )$ and dipole-dipole interaction potential $\Delta _{ij}(\omega
)$ are presented in Fig. \ref{ssp5}(a) and Fig. \ref{ssp5}(b) for atom A
located at $(0,-11/15,0)a$ and atom B located at $(0,11/15,0)a$. The insets
show the behaviors for frequency around the resonance frequency ($\omega
_{c}=0.3133(2\pi c/a)$). From Fig. \ref{ssp5}(a), we clearly see that $%
\Gamma _{ij}(\omega )$ oscillates remarkably when $\omega $ is far away from
$\omega _{c}$, which shows the powerful modulation ability of photonic
crystal for the electromagnetic eigen mode and the density of photonic
states. For $\omega =\omega _{c}$ (the inset), $\Gamma _{ij}(\omega
_{c})\approx 144\Gamma _{0}$ for relatively large atomic separation ($%
R\approx 0.46\lambda _{0}$), where $\Gamma _{0}$ is the largest $\Gamma
_{ij}(\omega _{0})$ for two dipoles located in vacuum for the atomic
separation $R=0$. This large $\Gamma _{ij}$ is attributed to the enhancement
of photon emission and reabsorbing rates, which can be roughly characterized
by the ratio of Q/V. Aimed for quantum computation and quantum information
processing \cite{Lukin1,Lukin2} where large $\Gamma _{ij}$ or large $\mathbf{%
\Delta }_{ij}$ is needed, we can increase the quality factor Q through
improving the nanocavity design. For $\omega \sim \omega _{c}$ (the inset), $%
\Gamma _{ij}(\omega )$ varies sharply and is sensitive to the frequency.
This implies that tuning $\omega _{c}$ of the cavity or $\omega _{i}$ of the
dipole can both help to control $\Gamma _{ij}$. Figure \ref{ssp5}(b) shows
similar properties for the dipole-dipole interaction potential. The inset
shows that repulsive or attractive potential can be got around the cavity
frequency.

\begin{figure}[htbp]
\includegraphics[scale=0.465]{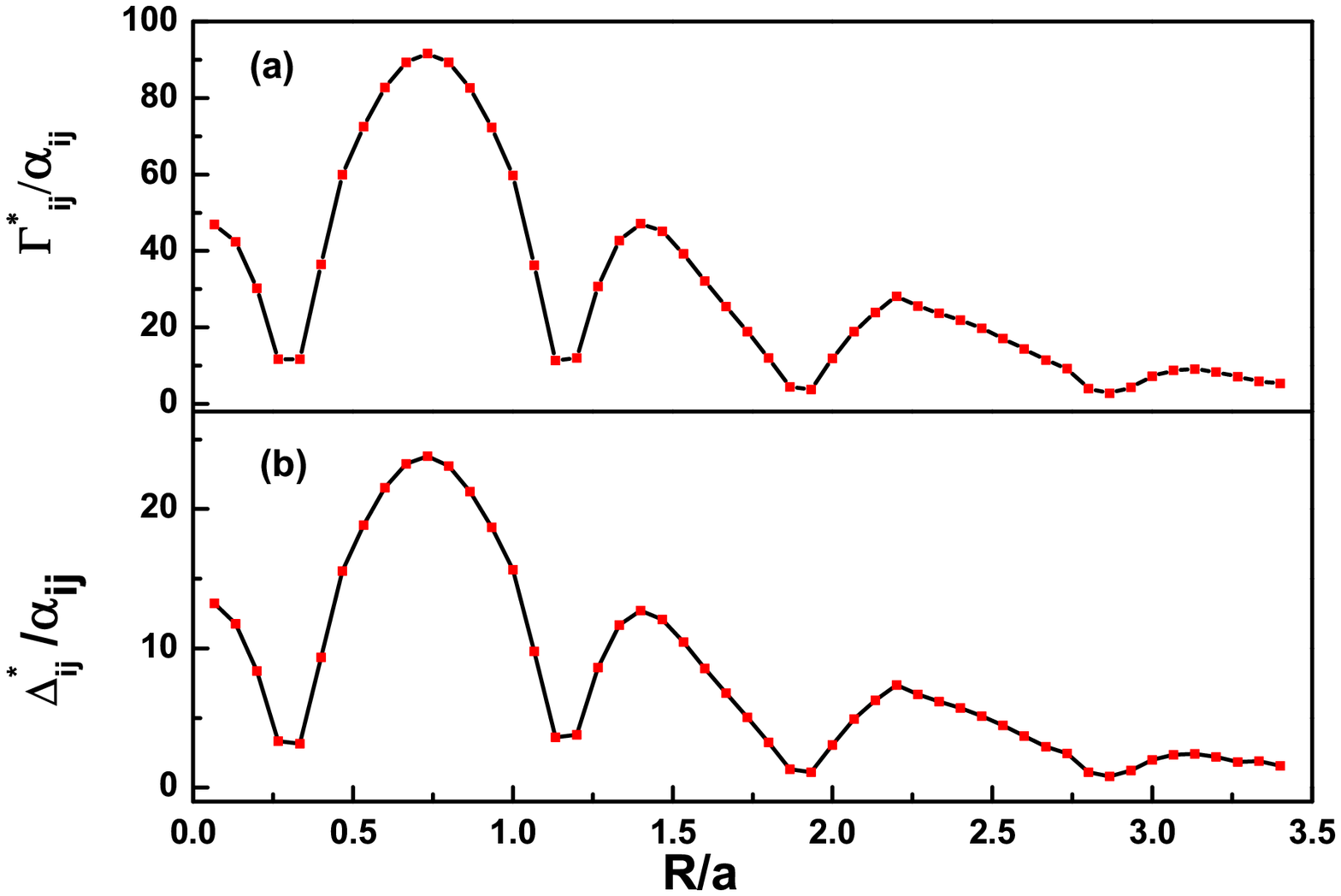}
\caption{(Color online). The position dependent characteristics for
cooperative decay parameters and dipole-dipole interaction potential. (a)
The absolute maximum cooperative decay parameters $\Gamma _{ij}^{\ast
}=\left\vert \Gamma _{ij}\right\vert $ and (b) the absolute maximum
dipole-dipole interaction potential $\mathbf{\Delta }_{ij}^{\ast
}=\left\vert \mathbf{\Delta }_{ij}\right\vert $ versus the interatomic
separation $R$ for atom A located at $(0,-11/15,0)a$ and atom B located at $%
(0,R-11a/15,0)$ in the PCs cavity. The maximum are calculated for $\protect%
\omega $ around $\protect\omega _{c}$. $\Gamma _{ij}^{\ast }$ and $\mathbf{%
\Delta }_{ij}^{\ast }$ are in unit of $\protect\alpha _{ij}$ ($\protect%
\alpha _{ij}\equiv \protect\sqrt{\protect\alpha _{i}}\protect\sqrt{\protect%
\alpha _{j}}2\protect\pi c/a$) and $\protect\omega $ is in unit of $2\protect%
\pi c/a$. The two transition dipoles are parallel and along the x axis.}
\label{ssp6}
\end{figure}

The position dependent characters have been investigated. Figure \ref{ssp6}%
(a) and \ref{ssp6}(b) show the absolute maximum value for $\omega $ around $%
\omega _{c}$ of cooperative decay parameters $\Gamma _{ij}^{\ast
}=\left\vert \Gamma _{ij}\right\vert $ and dipole-dipole interaction
potential $\mathbf{\Delta }_{ij}^{\ast }=$ $\left\vert \mathbf{\Delta }%
_{ij}\right\vert $ as a function of their separation R (atom A located at $%
(0,-11a/15,0)$ and atom B located at $(0,R-11a/15,0)$) are show where the
maximum are calculated for $\omega $ around $\omega _{c}$. $\Gamma
_{ij}^{\ast }$ and $\mathbf{\Delta }_{ij}^{\ast }$ are in unit of $\alpha
_{ij}$ ($\alpha _{ij}\equiv \sqrt{\alpha _{i}}\sqrt{\alpha _{j}}2\pi c/a$).
The $\omega $ is in unit of $2\pi c/a$. The maximum value $\Gamma
_{ij}^{\ast }$ and $\mathbf{\Delta }_{ij}^{\ast }$ varies in a similar way
except for the amplitude. The above phenomena can be understood as follows: $%
\Gamma _{ij}^{\ast }$ depends much on the electric field of the defect mode.
Equation (\ref{8}) shows that $\Gamma _{ij}^{\ast }$ changes in the same way
as the electric field of the cavity mode at the location of atom B. The
cavity mode along the y axis looks much the same as Fig. \ref{ssp6}(a) and
we did not show it here. Further, we clearly see that $\Gamma _{ij}^{\ast
}=\Gamma _{ii}^{\ast }$ and $\mathbf{\Delta }_{ij}^{\ast }=\mathbf{\Delta }%
_{ii}^{\ast }$ have been realized for atom A and atom B located at positions
with same electric field strength along the x axis.

\section{V. Conclusion}

In summary, we have proposed a new efficient and rigorous numerical method
to investigate dipole-dipole interaction in photonic crystal nanocavity. We
calculate the collective and individual radiation rates of classical dipoles
by directly solving Maxwell's equations in real space with a free-space
boundary condition. By using the result of two dipoles radiation rate minus
the sum of the two individual radiation rates, we get the cooperative decay
parameters and dipole-dipole interaction potential. Through the
self-consistent procedure (Eq. (\ref{4})), exact non-Markov results can be
got and both real and virtual photon effects have been taken into account.
Further, it can be applied for dipoles with different transition frequencies
in both weak and strong coupling regimes. Our investigation suggests that
this method works well for dipoles located in media with arbitrary shape and
may be generalized to calculate many dipoles interaction. Numerical
validation has been made in vacuum and planar nanocavity. Both results agree
very well with the analytic.

Applying this method to a simple photonic crystal nanocavity, it is found
that the cooperative decay parameters and the dipole-dipole interaction
potential strongly depend on the atomic position, the transition frequency,
quality factor and the cavity frequency. For two dipoles arranged at
positions with the same local coupling strength, the cooperative decay
parameters is equal to the local coupling strength. Large cooperative decay
parameters is achieved at the resonance frequency. Dipole-dipole interaction
potential changes continuously from attractive to repulsive case for
transition frequency varying in a domain of the cavity linewidth around the
resonance frequency. Larger value and sharper change of cooperative
parameters and dipole-dipole interaction can be obtained for higher quality
factor. Owing to the Photonic crystal nanostructure is one of the most
promising platform. It offers many advantages, including high quality
factor, small mode volume, integrability with waveguide and fixed position
of dipoles. In addition, many ingenious schemes for static and ultra fast
dynamic control of the quality factor, resonance frequency have been proposed. Based on these techniques for photonic crystal nanocavity and methods for tuning the transition frequency of quantum dots, our results provide some manipulative approaches for dipole-dipole interaction with potential application in various fields such as quantum computation and quantum information processing based on solid state nanocavity and quantum dot system.

\section{Acknowledgment}

This work was supported by NSFC under grants Nos. 10934010, 60978019, the
NKBRSFC under grants Nos. 2009CB930701, 2010CB922904, 2011CB921502,
2012CB821300, and NSFC-RGC under grants Nos. 11061160490 and
1386-N-HKU748/10.

\end{document}